# Using Statistical Moment Invariants and Entropy in Image Retrieval

Ismail I. Amr*, Mohamed Amin†, Passent El-Kafrawy†, and Amr M. Sauber†

*College of Computers and Informatics
Misr International University, Cairo, Egypt
†Faculty of science Department of Math and Computer Science
Menoufia University, Shebin-ElKom, Egypt

## Abstract

Although content-based image retrieval (CBIR) is not a new subject, it keeps attracting more and more attention, as the amount of images grow tremendously due to internet, inexpensive hardware and automation of image acquisition. One of the applications of CBIR is fetching images from a database. This paper presents a new method for automatic image retrieval using moment invariants and image entropy, our technique could be used to find semi or perfect matches based on query-by-example manner, experimental results demonstrate that the purposed technique is scalable and efficient.

## 1 Introduction

In many areas of commerce, government, academia, and hospitals, large collections of digital images are being created. Many of these collections are the product of digitizing existing collections of analogue photographs, diagrams, drawings, paintings, and prints. Usually, however, technologies related to archiving, retrieving, and editing images/video based on their content are still in their infancy, the only way of searching these collections was by keyword indexing, or simply by browsing. Digital image databases however, open the way to content-based searching. "Content-based" means that the search will analyze the actual contents of the image. The term content' in this context might refer to color, shape, texture, or any other information that can be derived from the image itself. Without the ability to examine image content, retrieval must rely on metadata such as captions or keywords, which may be laborious or expensive to produce.

What is desired is a similarity matching, independent of translation, rotation, and scale, between a given template (example) and images in the database. Consider the situation where a user wishes to retrieve all images containing cars, people, etc. in a large visual library. Being able to form queries in terms of sketches, structural descriptions, color, or texture, known as query-by-example (QBE), offers more flexibility over simple alphanumeric descriptions. This paper is organized as follows: section 2 provides the necessary background for CBIR. Section 3 defines the image segmentation technique using the Moments and Entropy. Section 4 explains the pro- posed model with the experiments conducted. Finally, section 5 concludes the paper and lists some further work.

### 1.1 Background

Among the approaches used in developing early image database management systems (IDMS) are textual encoding [1], logical records [2], and relational databases [3]. The descriptions, employed to convey the content of the image, were mostly alphanumeric. Furthermore, these were obtained manually or by utilizing simple image processing operations designed for the application at hand. Later generations of IDMS have been designed in an object-oriented environment [4], where image interpretation routines form the backbone of the system. However, queries still remain limited to a set of predetermined features that can be handled by the system. The reader is referred to [5] for a survey of IDMS.

Most recent systems reported in the literature for searching, organizing, and retrieving images based on their content include IBMs Query-by-Image-Content (QBIC) [6], MITs photo-book [7], the Trademark and Art Museum applications from ETL [8], Xenomania from the University of Michigan [9], and Multimedia/VOD test bed applications from the Columbia University [10]. IBMs QBIC is a system that translates visual information into numerical descriptors, which are then stored in a database. It can index and retrieve images based on average color, histogram color, texture, shape, and sketches. MITs photobook describes three content-based tools, utilizing principal component analysis, finite element modes, and the Wold transform to match appearances, shapes, and textures, respectively, from a database to a prototype at run time. Xenomania is a system for face image retrieval, which is based on QBE. Its embedded routines allow for segmentation and evaluation of objects based on domain knowledge, yielding feature values that can be utilized for similarity measures and image retrieval. The database management system of the Columbia University proposes integrated feature maps based on texture, color, and shape information for image indexing and query in transform domain.

Similarity-based searching in medical image databases has been addressed in [11]. A variety of shape representation and matching techniques are currently available, which are invariant to size, position, and/or orientation. They may be grouped as: (1) methods based on local features such as points, angles, line segments, curvature, etc. [12]; (2) template matching methods [13]; (3) transform coefficient based methods, including Fourier descriptors [14] or generalized Hough transform [15]; (4) methods using 3 modal and finite element analysis [16]; (5) methods based on geometric features, such as local and differential invariants [17]; and (6) methods using B-Splines or snakes for contour representation [18]. Comprehensive surveys of these methods can be found in [19].

### III. BACKGROUND

#### A. Image Segmentation

The shape representation method described here assumes that the object has been fully segmented from the original image, such that all pixels representing the objects shape have been identified as distinct from those pertaining to the rest of the image. In this paper, a local diffusive segmentation method [20] is used. There exist a wide variety of ways to achieve segmentation; however, it is not the subject of this paper. All contiguous pixels, which share a given point-based characteristic of the object or are surrounded by those that do, are considered as object pixels and those outside the included region, are considered as background. The result is a group of contiguous pixels, which collectively represent the object. The boundary pixels of the object are then extracted from the segmented object pixels by a simple iterative trace, around the outside of the object that continues until the starting point is reached. This trace produces a second group of pixels collectively representing the objects exterior contour [20].

#### B. Entropy

Entropy is a scalar value representing a statistical measure of randomness that can be used to characterize the texture of the input image. Entropy is defined as

$$S = -\sum_{i=1}^{\Omega} P_i \log_2(P_i)$$

The value of entropy is also an invariant that is neither affected by rotation nor scaling.

#### C. Moments

Region moment representations interpret a normalized gray level image function as a probability density of a 2D random variable. Properties of this random variable can be described using statistical characteristics - moments. A moment of order (p+q) is dependent on scaling, translation, rotation, and even on gray level transformations and is given by

$$m_{pq} = \int_{-\infty}^{\infty}\int_{-\infty}^{\infty} x^p y^q g(x,y) dx dy$$

The central moment

$$\mu_{pq} = \int_{-\infty}^{\infty}\int_{-\infty}^{\infty} (x-x_c)^p (y-y_c)^q g(x,y) dx dy$$

Let $f(x,y) = (x-x_c)^p (y-y_c)^q g(x,y)$

We use composite trapezoidal rule for evaluation the double integral

$$I_T = \int_{-\infty}^{M} \int_{-\infty}^{N} f(x,y) dx dy$$

$$I_T = \frac{hk}{4} \begin{bmatrix} \{f_{0,0} + f_{0,M} + 2(f_{0,1} + f_{0,2} + \ldots + f_{0,M-1})\} \\ +2\sum_{i=1}^{N-1} \{f_{i,0} + f_{i,M} + 2(f_{i,1} + f_{i,2} + \ldots + f_{i,M-1})\} \\ \{f_{N,0} + f_{N,M} + 2(f_{N,1} + f_{N,2} + \ldots + f_{N,M-1})\} \end{bmatrix}$$

Where $M$ and $N$ image size, $x_c$, $y_c$ are the co-ordinate of the region's centriod, $h = k = 1$ and

$$y_c = \frac{m_{01}}{m_{00}}, \quad x_c = \frac{m_{10}}{m_{00}}$$

The normalized un-scaled central moments

$$\vartheta_{pq} = \frac{\mu_{pq}}{(\mu_{00})^\lambda}$$

Where $\lambda = \frac{P+q}{2} + 1$

A less general form of invariance is given by seven rotation, translation, and scale invariant moment characteristics [31].

$$\phi_1 = \vartheta_{20} + \vartheta_{02}$$
$$\phi_2 = (\vartheta_{20} - \vartheta_{02}) + 4\vartheta_{11}^2$$
$$\phi_3 = (\vartheta_{20} - 3\vartheta_{12})^2 + (3\vartheta_{21} - \vartheta_{03})^2$$
$$\phi_4 = (\vartheta_{30} - \vartheta_{12})^2 + (\vartheta_{21} - \vartheta_{03})^2$$
$$\phi_5 = \begin{bmatrix} (\vartheta_{30} - 3\vartheta_{12})(\vartheta_{30} + \vartheta_{12})\left[(\vartheta_{30} + \vartheta_{12})^2 - 3(\vartheta_{21} + \vartheta_{03})^2\right] \\ +(3\vartheta_{21} - \vartheta_{03})(\vartheta_{21} + \vartheta_{03})\left[3(\vartheta_{30} - \vartheta_{21})^2 - (\vartheta_{21} + \vartheta_{03})^2\right] \end{bmatrix}$$

$$\phi_6 = \begin{bmatrix} (\vartheta_{20} - \vartheta_{02})\left[(\vartheta_{30} + \vartheta_{12})^2 - (\vartheta_{21} + \vartheta_{03})^2\right] \\ +4\vartheta_{11}(\vartheta_{30} + \vartheta_{12})(\vartheta_{21} + \vartheta_{03}) \end{bmatrix}$$

$$\phi_7 = \begin{bmatrix} (3\vartheta_{21} - \vartheta_{03})(\vartheta_{30} + \vartheta_{12})\left[(\vartheta_{30} + \vartheta_{12})^2 - 3(\vartheta_{21} + \vartheta_{03})^2\right] \\ -(\vartheta_{30} - 3\vartheta_{12})(\vartheta_{21} + \vartheta_{03})\left[3(\vartheta_{30} + \vartheta_{21})^2 - (\vartheta_{21} + \vartheta_{03})^2\right] \end{bmatrix}$$

## 2 The Proposed Technique

Our proposed method consists of two parts: region selection and shape matching. In the first part, the image is partitioned into disjoint, connected regions. The second part, the shape of each potential object is tested to determine whether it matches one from a set of given template. To this effect, we use image Moments and Entropy. Although many techniques [21]–[23] use moment invariants to overcome rotation and scaling, we extend this technique by using both moments and entropy to support two level filtering which is more appropriate in an image database case. In this paper we focus on part two and part one may be implemented as shown in section III-A.

A. Implementation

Images, as figure 1, are stored in the database after its subdivided into distinct sub-images as stated previously. For each sub-image stored we compute the entropy and moments, that is indexed accordingly.

Each time a template image is requested to find its match in the database, the following steps are implemented. First we compute the entropy and moments for the image that is being searched for. Second, the computed entropy is used to filter the images in the database. At last, the computed moments are used to determine and fetch the most matching images in the filtered subset.

B. Experimental Results

The experiments run on an image database that contains one hundred images. Figure 1 shows all the images, where each of these sub-images are stored independently in the database. We picked some images in the database scaled, rotated, or scaled and rotated and used as testing templates. Figure 2 and 3 show multiple experimental results that demonstrate the performance of our technique, each experiment contains a template (example) and a result set containing the matches. All the results are promising except when the template is scaled larger than double the size of the original image (i.e. image in the database).

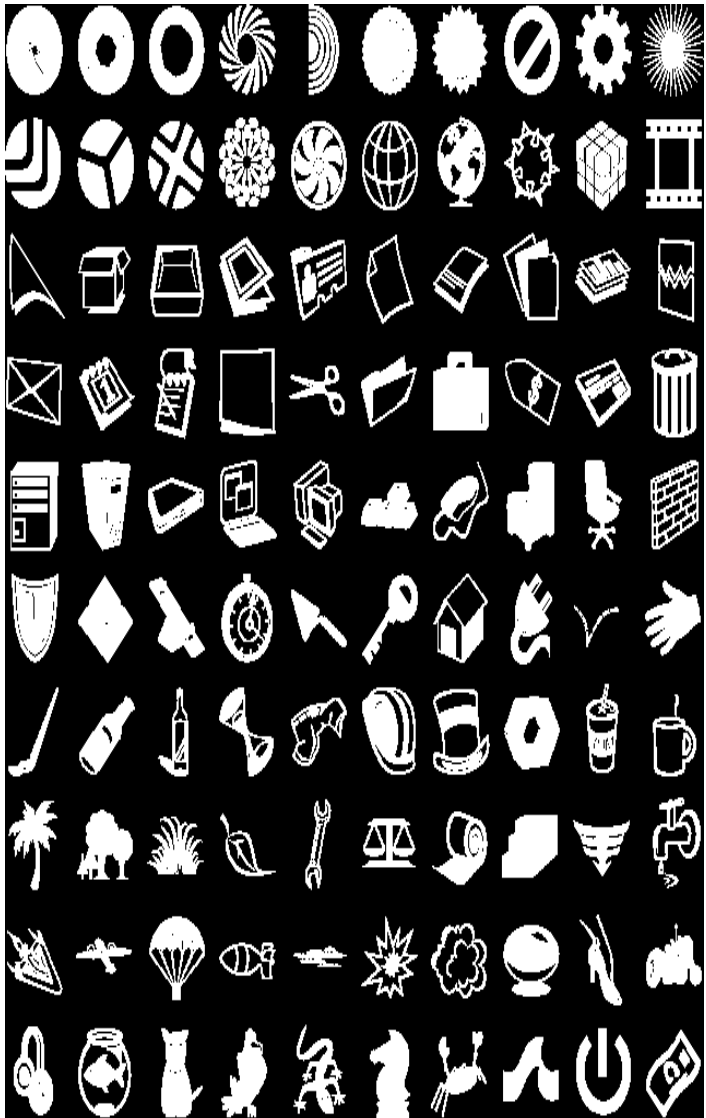

Figure 1 Images in database

Experiments

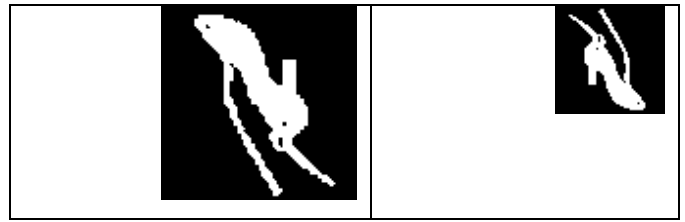

| Template | Result |
|---|---|
|  |  |
|  |  |

## Conclusions

This paper presents a method for automatic CBIR based on query-by-example by region-based shape matching. The proposed method consists of two parts: region selection and shape matching. Each image is segmented to a set of sub images by using a local diffusive segmentation method. Consequently, the image entropy is computed and used to narrow the search space. Later, the moment invariants of the image are matched, which are independent to translation, scale, rotation and contrast, to every sub image and to the template. Retrieval is performed by returning those sub images whose invariant moments are most similar to the ones of a given query image.

In the future we would like to find more image characteristics that can be used to query images in the database. Our focus now is to explore other characteristics that can be used for highly scaled images. Moreover, we will research colored images and develop new techniques to handle color.

ACKNOWLEDGMENT

The author would like to thank the valuable comments and encouragement received from colleagues in Menoufia University.